# A Monitoring Language for Run Time and Post-Mortem Behavior Analysis and Visualization

Mikhail Auguston[1], Clinton Jeffery[2], Scott Underwood[2]

[1]Department of Computer Science, Naval Postgraduate School
auguston@cs.nps.navy.mil

[2]Department of Computer Science, New Mexico State University
{ jeffery, sunderwo}@cs.nmsu.edu

**Abstract**

*UFO is a new implementation of FORMAN, a declarative monitoring language, in which rules are compiled into execution monitors that run on a virtual machine supported by the Alamo monitor architecture. FORMAN's event trace model uses precedence and inclusion relations to define a DAG structure that abstracts execution behavior. Compiling FORMAN rules into hybrid run-time/post-mortem monitors gives substantial speed and size improvements over pure post-mortem analyzers that operate on the event trace DAG. The UFO optimizing compiler generates code that computes, at run-time when possible, the minimal projection of the DAG necessary for a given set of assertions. UFO enables fully automatic execution monitoring of realistic size programs. The approach is non-intrusive with respect to program source code. The ability to compile suites of debugging and program visualization rules into efficient monitors, and apply them generically to different programs, will enable long-overdue breakthroughs in program debugging automation.*

## 1. Motivation

Debugging is one of the most challenging and least developed areas of software engineering. Debugging activities include queries regarding many aspects of target program behavior: sequences of steps performed, histories of variable values, function call hierarchies, checking of pre- and post-conditions at specific points, and validating other assertions about program execution. Performance testing and debugging involves a variety of profiles and time measurements.

We are building automatic debugging and program visualization tools based on precise program execution behavior models that enable us to employ a systematic approach. Our program behavior models are based on events and event traces [1][2][3]. Any detectable action performed during a target program's run time is an *event*. For instance, expression evaluations, statement executions, and procedure calls are all examples of events. An event has a beginning, an end, and some duration; it occupies a time interval during program execution. This leads to the introduction of two basic binary relations on events: partial ordering and inclusion. Those relations are determined by target language syntax and semantics, e.g. two statement execution events may be ordered, or an expression evaluation event may occur inside a statement execution event. The set of events produced at run time, together with ordering and inclusion relations, is called an *event trace* and represents a model of program behavior. An event trace forms a directed acyclic graph (DAG) with two types of edges corresponding to the basic relations.

In our approach, the term debugging automation refers to a computation over an event trace. *Program execution monitors* are programs that load and execute a target program, obtain events at run-time, and perform computations over the event trace. Computations are performed during execution, post-mortem, or in any mixture of both times.

The language UFO[4] (from Unicon-FORMAN) integrates the experience accumulated in the FORMAN [1] language and the Alamo monitoring architecture [17] to provide a complete solution for development of an extensive suite of automatic debugging tools. UFO is an implementation of FORMAN for debugging programs written in the Unicon and Icon programming languages [18][12].

The reasons visualization is needed in this domain should be obvious: debugging is often performed at great cost in a depressing scenario in which the programmer (prospective user of visualization tools) probably does not understand what is wrong with the program, may not be one of the persons who wrote the code, and does not know where to look. The UFO approach provides for automatic generation of runtime visualization tools, based on declarative specifications of behaviors of interest. This approach will enable developers to use visualization in an exploratory fashion to discover unknown behaviors.

The suggested approach is nondestructive with respect to the target code. Text of assertions, visualization rules,

and other rules is separated from the target code. Instrumentation of the target code is done automatically.

## 2. Unicon and Alamo

The Unicon language and the Alamo monitoring architecture provide the underlying research framework for the implementation of UFO. Unicon is an imperative, goal-directed, object-oriented superset of the Icon programming language. Unicon's syntax is similar to Pascal or Java, while its semantics are higher level, featuring built-in backtracking, heterogeneous data structures and string scanning facilities. Icon has influenced scripting languages such as Python. Unicon is Icon's direct descendant; it extends Icon's reach with object-orientation, packages, a rich system interface with high level graphics, networking, and database facilities.

The reference implementation of Unicon is a virtual machine. Virtual machines (VMs) are attractive to language implementers because they provide portability and allow vastly simpler implementation of very high level language features such as backtracking.

The reasons VMs are attractive to language implementers also make them ideal for developing debugging tools. VMs provide an appropriate level of abstraction for developing behavior models to describe program executions in a processor independent manner, as illustrated by the JPAX tool described in the related work section below [14]. VMs provide easy access to program state and control flow. Automatic instrumentation on multiple semantic levels is greatly simplified via the use of a VM. This potential is exploited in the Unicon VM by a framework that implements the Alamo monitor architecture. Event instrumentation and processing support are an integral part of the VM.

The Alamo Unicon framework is summarized in Figure 1. Execution monitors (EM) and the target program (TP) execute as (sets of) coroutines with separate stacks and heaps inside a common VM. The VM is instrumented with approximately 150 kinds of atomic events, each one reporting a <code,value> pair. EMs specify categories of events by supplying an event mask when they activate the TP by coroutine switch (**1**). The TP executes up to an event of interest.

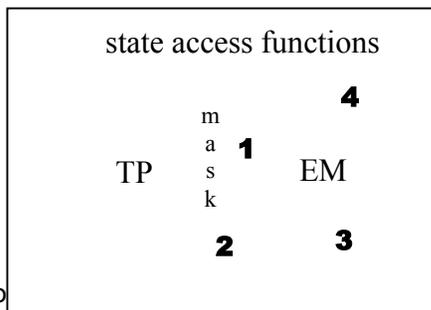

Figure 1. Alamo architecture within the Unicon VM.

The event mask is used by the VM for instrumentation selection and control (**2**). Event reports during TP execution are coroutine context switches from the VM runtime system back to the execution monitor (**3**). In addition to the information reported explicitly for the event, the EM can directly access arbitrary variable values and state information from the TP via state access functions (**4**).

Monitors are written independently from the target program, and can be applied to any target program without recompiling the monitor or target program. Monitors dynamically load target programs, and can easily query the state of arbitrary variables at each event report. Multiple monitors can monitor a program execution, under the direction of a monitor coordinator.

Alamo reduces the difficulty of writing execution monitors to be the same as writing other types of application programs. UFO moves beyond Alamo to reduce the difficulty of writing automatic debuggers and visualizers to the task of specifying generic assertions about program behavior. UFO's FORMAN language is described in Section 4 below, but first it is necessary to present the underlying behavior model.

## 3. An Event Grammar for Unicon

FORMAN uses an event grammar to provide a model of program run time behavior. Monitors do not have to parse events using the grammar, since event detection is part of virtual machine and UFO runtime system functionality. Monitors implement computations over event traces supplied by the virtual machine.

A FORMAN event is an abstraction of a group of underlying Alamo (Unicon VM) events that form a meaningful action of interest at run time. FORMAN events have event type, duration, and other attributes. Event type forms the primary behavior pattern matching construct in the model; attributes refine search queries.

The FORMAN events, attributes, and grammar summarized in this section provides a "lightweight" semantics of the Unicon programming language tailored for specification of debugging activities. The underlying Alamo event monitoring mechanism is described in [17].

**Universal attributes** are found in every event. They are frequently used to narrow assertions down to a particular domain (function, variable, value) of interest. Some of these attributes are much easier to obtain than others, and affect the optimizations that can be performed when generating monitor code; see Section 6 for details.

- source_text: in canonical form (i.e. with redundant spaces eliminated, etc.)
- line_num, col_num: source text locations
- time_at_end, time_at_begin, duration: timing attributes
- eval_at_begin ( Unicon-expression),
- eval_at_end ( Unicon-expression): these attributes provide access to the program states
- prev_path, following_path: set of events completed before (after) the event

The event types, and type-specific attributes they provide, are summarized in the table below.

| Event Type | Description | Attributes |
|---|---|---|
| prog_ex | whole program execution | |
| expr_eval | expression evaluation | value, operator, type, failure_p |
| func_call | function call | func_name, paramlist |
| input, output | I/O | file |
| variable | variable reference | |
| literal | reference to a constant value | |
| lhp | lefthand part, assignment | address |
| rhp | righthand part, assignment | |
| clause | then-, else-, or case branch execution | |
| test | test evaluation | |
| iteration | loop iteration | |

Event types form the following hierarchy.

```
expr_eval :
   ├── func_call : input, output
   └── variable, literal, clause, iteration, test, lhp, rhp
```

Subtypes inherit attributes from the parent type. Expression evaluation is the central action during Unicon program execution, this explains why the expr_eval event is on the top of the hierarchy.

The UFO *event grammar* for Unicon is a set of axioms describing the structure of event traces with respect to the two basic relations: inclusion and precedence. It is not intended to be used for parsing event traces. The event types presented above are the symbols used in this description. The grammar is one possible abstraction of Unicon semantics; other event grammars with far more (or less) detail might be used. The event grammar limits what kinds of bugs can be detected. The grammar uses the following notation:

| Notation | Meaning |
|---|---|
| A :: (B C) | B precedes A, A includes B and C |
| A* | Zero or more A's under precedence |
| A+ | One or more A's under precedence |
| A \| B | Either A or B; alternative |
| { A , B } | Set; A and B have no precedence |

prog_ex::   ( expr_eval *)
expr_eval::( ( expr_eval ) |               *unary op*
            ( expr_eval expr_eval ) |      *binary op*
            ( expr_eval+ ) |
            ( test clause ) |              *conditional/ case expressions*
            ( iteration * ) |              *loops*
            ( { lhp, rhp} )                *assignment lhp and rhp are not ordered, beginning of lhp precedes rhp, and end of lhp follows rhp*
           )
iteration::   ( test expr_eval*) | ( expr_eval* test ) | ( expr_eval * )

Execution of a Unicon program produces a set of events (an *event trace*) organized by precedence and inclusion into a DAG. The structure of the event trace (event types, precedence and inclusion of events) is constrained by the event grammar axioms above. The event trace models Unicon program behavior and provides a basis to define different kinds of debugging activities (assertion checking, debugging queries, profiles, debugging rules, behavior visualization) as appropriate computations over the event traces.

## 4. FORMAN

Alamo allows efficient monitors to be constructed in Unicon, but using a special-purpose language such as FORMAN, with the rich behavior model described in the preceding section, has compelling advantages. UFO rules are generally an order of magnitude smaller (in terms of lines of source code) than the equivalent monitors written in Unicon, depending on the type of quantifiers and aggregate operations used in the FORMAN rule.

FORMAN's support for computations over event traces centers around the notions of event patterns, aggregate operations, and quantifiers over events.

The simplest event pattern consists of a single event type and matches successfully an event of this type or an event of a subtype of this type. Event patterns may include event attributes and other event patterns to specify the context of an event under consideration. For example, the event pattern

  E: expr_eval  & E.operator == ":="

matches an assignment event. Temporary variable E provides an access to the events under consideration within the pattern.

The following example demonstrates the use of an aggregate operation.
  CARD[ A: func_call &
              A.func_name == "read" FROM prog_ex ]
yields a number of events satisfying a given event pattern, collected from the whole execution history. Expression […] is a list constructor and CARD is an abbreviation for a reduction of '+' operation over the more general list constructor (SUM can be used instead of +/ as well):
  +/[A: func_call & A.func_name == "read"
      FROM prog_ex APPLY 1 ]
Quantifiers are abbreviations for reductions of Boolean operations OR and AND. For instance,
  FOREACH Pattern FROM event_set Boolean_expr
is an abbreviation for
  AND/[Pattern FROM event_set APPLY Boolean_expr ]
Debugging rules in FORMAN usually have the form:
*Quantified_expression*
        WHEN SUCCEEDS   *SAY-clauses*
         WHEN FAILS   *SAY-clauses*
The Quantified-expression is optional and defaults to TRUE. The execution of FORMAN programs relies on the Unicon monitors embedded in a virtual machine environment. Section 5 below describes the architecture of the UFO compiler and runtime system, which translates FORMAN to Unicon VM monitor code.

### 4.1. Application-Specific Analyses

This section presents formalizations of typical debugging rules. UFO supports traditional precondition checking, or **print** statement insertion, without any modification of the target program source code. This is especially valuable when the precondition check or print statement is needed in many locations scattered throughout the code.

**Example #1: Tracing**. Probably the most common debugging method is to insert output statements to generate trace files, log files, and so forth. It is possible to request evaluation of arbitrary Unicon expressions at the beginning or at the end of events. The virtual machine evaluates these expressions at the indicated time moments.

  FOREACH   A: func_call &
          A.func_name == "my_func"
      FROM prog_ex
   A.value_at_begin(
       write("entering my_func, value of X is:", X) ) AND
   A.value_at_end(
     write("leaving my_func, value of X is:", X) )

This debugging rule causes calls to **write()** to be evaluated at selected points at run time, just before and after each occurrence of event A. It is even simpler to write a rule that applies to all functions, where the condition on A.func_name can be omitted.

**Example #2: Profiling**. A myriad of tools are based on a premise of accumulating the number of times a behavior occurs, or the amount of time spent in a particular activity or section of code. The following debugging rule illustrates such computations over the event trace.

  SAY( "Total number of read() statements: "
        CARD[ r: input & r.filename == "xx.in"
              FROM prog_ex ]
       "Elapsed time for read operations is: "
       SUM [ r: input & r.filename == "xx.in"
              FROM prog_ex   APPLY r.duration] )

**Example #3: Pre- and Post- Conditions**. Typical use of assertions includes checking pre- and post-conditions of function calls.

  FOREACH A:func_call & A.func_name=="sqrt"
       FROM prog_ex
   A.paramlist[1] >=0 AND
   abs(A.value*A.value-A.paramlist[1]) < epsilon
  WHEN FAILS SAY("bad sqrt(" A.paramlist[1]
                                            ") yields " A.value)

### 4.2. Generic Bug Descriptions

Another prospect is the development of a suite of generic automated debugging tools that can be used on any Unicon program. UFO provides a level of abstraction sufficient for specifying typical bugs and debugging rules.

**Example #4: Detecting Use of Un-initialized Variables**. Reading an un-initialized variable is permissible in Unicon, but often leads to errors. In this debugging rule all variables in the target program are checked to ensure that they are initialized before they are used.

FOREACH V: variable  FROM prog_ex
           FIND D: lhp FROM V.prev_path

```
      D.source_text == V.source_text
WHEN FAILS SAY( " uninitialized variable "
      V.source_text)
```

**Example #5: Empty Pops**. Removing an element from an empty list is typical of expressions that fail silently in Unicon. While this can be convenient, it can also be a source of difficult to detect logic errors. This assertion assures that items are not removed from empty lists.

```
FOREACH  a: func_call &
     a.func_name == "pop" AND
     a.value_at_begin( *a.paramlist[1] == 0)
  SAY("Popping from empty list at event " a)
```

## 5. Examples of visualization rules

This section presents examples of typical visualization rules supported by our prototype. The graphics are simple; the point is how the information to be visualized is obtained. Future work will automate the construction of declarative visualization specifications via a graphical front-end, and integrate them into a debugger.

**Example #6: Point Plotter**. One of the simplest and yet most general visualization methods is to show a point plot of values. Distinct point sets on the same graph are distinguished using color, point size, or shape. The x axis is often used to denote logical sequence of events, or CPU or wall-clock time. The following FORMAN example specifies a point plot of several variables within a binary search program. The complete syntax of rules is given in the Appendix 1. As in Example #1, the use of FORMAN VALUE_AT_END() clauses are used to mix in Unicon expressions evaluated at appropriate points at run time.

```
SHOW POINT_PLOT (
 TITLE "History of: bottom (magenta),
     mid (green), top(blue)"
 WINDOW_SIZE 500 500
  X_SIDEBAR
   SCALING  linear
   TICKNUM 6
   MOVING  fixed
   INTERVAL_BEGIN  0
   INTERVAL_END 30
   TEXT_LABEL "Time counter"
  Y_SIDEBAR
   TICKNUM 12
   MOVING rescale
   INTERVAL_BEGIN  0
   INTERVAL_END 12
   TEXT_LABEL "Values"
 SOURCE a: expr_eval &
    a.LINE_NUM == 10
 X_AXIS ORD(a)
 SET
  COLOR green
  Y_AXIS a.VALUE_AT_END( mid )
  connected
 SET
  COLOR blue
  Y_AXIS a.VALUE_AT_END( top )
  connected
 SET
  COLOR magenta
  Y_AXIS a.VALUE_AT_END( bottom )
  connected
)
```

This rule plots the values of variables mid, top, and bottom in order of appearance of events defined in the SOURCE event pattern for the following toy Unicon program bsearch.icn, which does a binary search.

```
# binary search example
procedure main ()
 bottom := 1
 top    := 10
 A := [3,6,19,43,58,112,367,390,427,471]
 target := 367
 found := 0
 mid := (top + bottom) / 2

 while bottom < top do {
       #**** ERROR: must be <=
  if target < A[mid] then
    top := mid
  else bottom := mid +1
  mid := (top + bottom) / 2
  }
 if A[mid] = target then found := 1
 write("found= ", found)
end
```

Since the program contains a bug, the mid variable converges on the wrong value, 8. At the same time, we can visualize the expected property of the three values (mid is positioned between top and bottom).

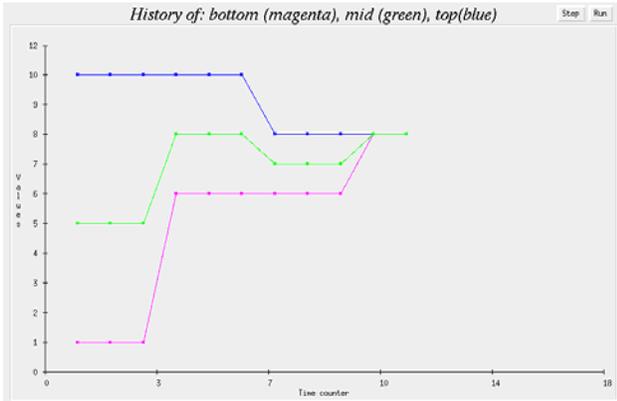

The following diagram, for the corrected program, demonstrates proper convergence to the correct value, 7. In arbitrary programs, visualizing many (or all) variable behaviors together in such a diagram may reveal unanticipated correlations and suggest places to examine in further detail. Other candidates where a pointplot may be of use include plotting history of function return values, history of expressions and intermediate results.

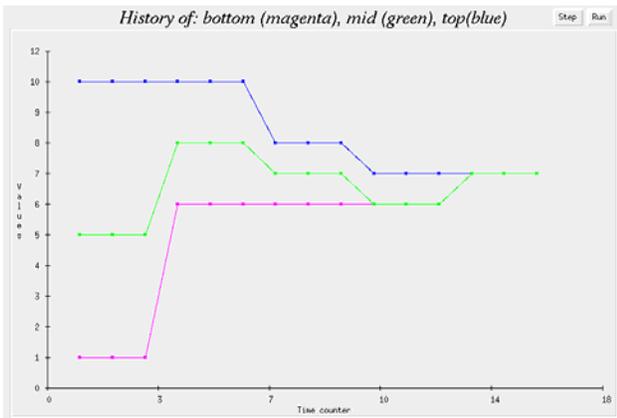

Point plots are very general; for example, the x axis may be used for arbitrary $2^{nd}$-variable dependencies, so that the point represents a pair of variable values. As another example, each point in the plot may represent one loop within the program; the y axis might be the number of times the loop was executed, and the x axis the number of iterations within the loop (scaled totals or averages).

**Example #7: Bar Chart**.

A bar chart is useful when the x axis is populated with nonnumeric values, e.g. function names. The following rule produces at the run time a bar chart when each bar corresponds to a function call events with a unique function name. The y axis is defined by an expression ORD(a). In the context of SHOW BAR_CHART rule the semantics for y axis expression requires to accumulate the value for each x axis value separately.

```
SHOW BAR_CHART (
  TITLE "Function call profile"
    X_SIDEBAR
        SCALING  linear
        TICKNUM      10
        MOVING         fixed
        INTERVAL_BEGIN  0
        INTERVAL_END 400
    Y_SIDEBAR
        TICKNUM      7
        MOVING rescale
    SOURCE  a: func_call
    X_AXIS  a.FUNC_NAME
    SET
        FOREGROUND_COLOR  yellow
        Y_AXIS  ORD(a)
    SET
        FOREGROUND_COLOR  blue
        Y_AXIS  ADD(a.DURATION) )
```

The semantics of source for the visualization is the same as for the previous example: the event pattern

a: func_call

determines the event flow feeding the visualization rule. The values for y axis are calculated and stored separately for each distinct x axis value. As a result this rule will accumulate and plot a profile and total duration (in milliseconds) for each function. The resulting image produced by the visualization tool for a sample word concordance program concord.icn looks like:

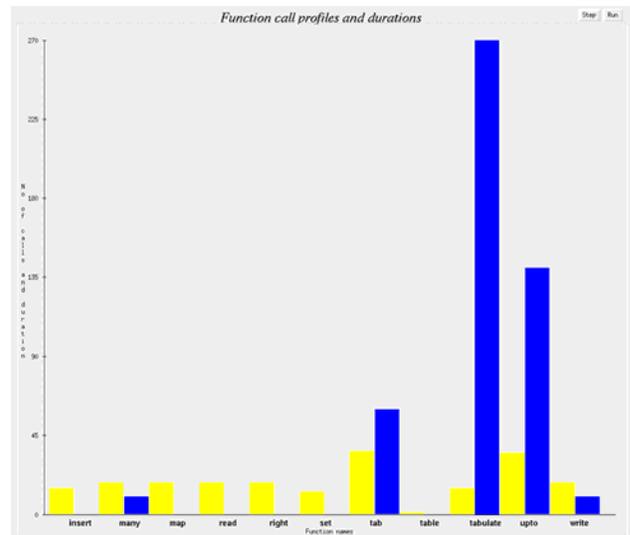

This is an example of a generic visualization rule that may be applied to any target program, and may be useful for performance debugging. A library of such rules will relieve most users from the challenges of writing visualization rules from scratch.

## 6. Implementation Issues

This section describes issues that have arisen during the implementation of UFO. The most important of these issues is the translation model by which FORMAN assertions are compiled down to Unicon Alamo monitors. Debugging activities are written as if they have the complete post-mortem event trace, the DAG with events, event attributes, and precedence and containment relations, available for processing. This generality is extremely powerful; however, for most practical uses we have seen, assertions can be compiled down into monitors that execute entirely at runtime. Runtime monitoring saves enormously on memory and I/O requirements and is the key to practical implementation. For those assertions that require post-mortem analysis, the UFO runtime system will compute a projection of the execution DAG necessary to perform the analysis.

### 6.1. Translation Examples

The UFO compiler generates Alamo Unicon monitors from FORMAN rules. Each FORMAN statement is translated into a combination of initialization, run-time, and post-mortem code. Monitors are executed as coroutines with the Unicon target program. The following examples give a flavor of the run time architecture of monitors generated from the UFO high level rules.

**Implementation of Example #1: Tracing.** This is a handwritten Unicon Alamo monitor for Example #1 above; the machine generated monitor is equivalent but longer and is omitted for space reasons. A single FOREACH quantifier is quite typical of many UFO debugging actions and allows computation to be performed entirely at runtime. The events being counted and values being accumulated are used to construct an *event mask* in the initialization code that defines the Alamo events that will be monitored.

The initialization for this monitor calls Alamo's EvInit() function to dynamically load the target program, and then computes an Alamo event mask, based on the FORMAN func_call event, the mask in this case requires the underlying Alamo procedure entry and exit events (E_Pcall, E_Pret and so forth).

The monitor's event processing loop calls Alamo's EvGet() function to obtain events, and implements the filter based on procedure name within an if-expression. The Unicon code blocks containing write() expressions are inserted directly into the event loop for the relevant events. The complete monitor is:

```
$include "evdefs.icn"
link evinit
procedure main(av)
  EvInit(av) | stop("can't monitor ", av[1])
  mask := E_Pcall ++ E_Pret ++ E_Pfail ++ E_Prem
  while EvGet(mask) do {
    if &eventcode == E_Pcall &
      &eventvalue === my_func then {
        write("entering my_func, value of X is:", X)
      }
    if &eventcode == (E_Pret | E_Pfail | E_Prem) &
      &eventvalue === my_func then {
        write("leaving my_func, value of X is:", X)
      }
  }
end
```

**Implementation of Example #2: Profiler.** This is the Unicon monitor corresponding to Example #2 above. This is another typical situation, which involves an aggregate operation and selection of events according to a given pattern. The SAY expression is implemented by a call to write(); it must be performed post-mortem since it uses parameters whose values are constructed during the entire program execution. CARD denotes a counter, while SUM denotes an accumulator +/; both require a variable that is initialized to zero. The event subtypes and constraints are used to generate additional conditional code in the body of the event processing loop. Lastly, some attributes such as r.duration require additional events and measurements besides the initial triggering event. In the case of r.duration, a time measurement between the function call and its return is needed.

```
$include "evdefs.icn"
link evinit
procedure main(av)
  EvInit(av) | stop("can't monitor ", av[1])
  cardreads := 0
  mask := E_Fcall
  sumreadtime := 0
  while EvGet(mask) do {
    if &eventcode == E_Fcall &
      &eventvalue === (read|reads) then
        cardreads +:= 1
    if &eventcode == E_Fcall &
      &eventvalue === (read|reads) then {
        thiscall := &time
        EvGet(E_Ffail++E_Fret)
        sumreadtime +:= &time - thiscall
      }
  }
  write("Total number of read() statements: ",
```

```
            cardreads, "\n",
         "Elapsed time for read operations is: ",
            sumreadtime)
   end
```

## 6.2. Basic Generation Templates

   The preceding handwritten example monitors use a single main loop that implements traditional event-driven processing. Monitors generated by the UFO compiler reduce complex assertions to this same single event loop. Keeping event detection in a single loop allows uniform processing of multiple event types used by multiple monitors. The code generated by the UFO compiler integrates event detection, attribute collection, and aggregate operation accumulation in the main event loop.

   Assertions in UFO may use nested quantifiers implying two nested loops, so code generation addresses this issue by flattening the main loop structure, and postponing assertion processing until required information is available. A hybrid code generation strategy performs runtime processing whenever possible, delaying analyses until post-mortem time when necessary. Different assertions require different degrees of trace projection storage; code responsible for trace projection collection is also arranged within the main loop.

   Each UFO rule falls in one of the following categories which determines its code generation template in the current implementation. We have not found a use for assertions requiring more than two nested quantifiers.

**I) Rules with a single quantifier or aggregate operation.**

   The only possible range for event selection is the whole trace (the event prog_ex). These rules can be evaluated entirely at run time without accumulating trace projections; examples in Section 4.1 illustrate this case.

   The following templates are used either for rules with two nested quantifiers or one quantifier and one aggregate operation whose range may depend on the metavariable bound by the first quantifier.

**II) Rules with two nested quantifiers of the form**

```
   Quantifier A: Pattern_A
        Quantifier B: Pattern_B FROM A
             Body
 The monitor's main loop follows the pattern:
    Main Loop
         Maintain stack of nested A events
         Accumulate events B in a B-list
         If end of event A
           Loop over B-list
                Do Body
         Endif
```
         If stack of A is empty
              Destroy B-list
    End of Main Loop

This requires accumulation of a trace projection for B-events and incurs a mild overhead at run time.

**III) Rules with two nested quantifiers of the form**

```
   Quantifier A: Pattern_A
        Quantifier B: Pattern_B FROM A .prev_path
             Body
The monitor's main loop follows the pattern:
    Main Loop
         Maintain stack of nested A events
         Accumulate events B in a B-list
         If end of event A
           Loop over B-list
                If B precedes A
                     Do Body
         Endif
    End of Main Loop
```

This requires accumulation of a trace projection for B-events and may cause a heavy overhead at the run time. Notice that B-list can not be deleted till the end of session.

**IV) Rules with two nested quantifiers of the form**

```
   Quantifier A: Pattern_A
        Quantifier B: Pattern_B
             FROM A .following _path
             (or FROM prog_ex)  Body
The monitor's main loop follows the pattern:
    Main Loop
         Accumulate events A in A-list
         Accumulate events B in B-list
    End of Main Loop
    # Postmortem Loop
    Loop over A-list
         Loop over B-list
              Do Body
    End of Postmortem Loop
```

   This requires accumulation of trace projections for both A and B events and may cause a very heavy overhead at run time.

## 6.3. Optimization Issues

   The UFO approach combines an optimizing compiler for monitoring code with efficient run-time event detection and reporting. Since we know at compile time all necessary event types and attributes required for a given UFO rule, the generated Unicon monitor can be very selective about the behavior that it observes.

For certain kinds of UFO constructs, such as nested quantifiers, the monitor must accumulate a sizable projection of the complete event trace and postpone corresponding computations until all required information is available. The presence of the previous_path and following_path attributes in UFO assertions facilitates this kind of optimization.

For further optimization, especially in the case of programs containing a significant number of modules, the following FORMAN construct limits event processing to events generated within the bodies of functions F1, F2, … , Fn.

    WITHIN F1, F2, … , Fn DO
        Rules
    END_WITHIN

This provides for monitoring only selected segments of the event trace.

Unicon expressions included in the value_at_begin and value_at_end attributes are evaluated at run time.

Some other optimizations implemented in this version are:

- only attributes explicitly used in the UFO rule are collected in the generated monitor;
- an efficient mechanism for event trace projection management, which disposes from the stored trace projection those events that will not be used after a certain time (for example, see Category II);
- both event types and context conditions are used to filter events for the processing.

UFO's goal of practical application to real-sized programs has motivated several improvements to the already carefully-tuned Alamo instrumentation of the Unicon virtual machine. We are working on additional optimizations.

## 7. Results of Sample Assertion Execution

The following table provides results from the execution of assertions written in UFO on a realistic target program. In Table 1, the target program is igrep.icn, an 1,100 line program that emulates UNIX egrep. By passing text files of varying length to igrep, we demonstrate the performance of UFO on varying numbers of events. Tests were performed on a SunBlade 1000 running a Solaris operating system with a 700MHz CPU and 512MB of RAM. The results reported for each input size are number of events generated by the VM and execution time averaged over several runs. Execution time is reported in the manner returned by the UNIX time command: minutes:seconds.tenths The top row of Table 1 indicates the number of lines of text within the input file passed to the target program. In order to provide a base case from which the overall cost of monitoring can be determined, the next row contains the results of target program execution without monitoring. Each target program/input file combination was monitored by 8 different assertions corresponding to the basic generation templates discussed in section 5.2.

Cases 1-4 are examples of a Category I generation template. Case 5 is a Category II assertion. Case 6 is a Category III assertion. It uses PREV_PATH and accumulates one trace projection over a part of the program execution. Processing of this projection is integrated in the main monitoring loop, and performed when the necessary events are all available. Cases 7 and 8 contain nested quantifiers that belong to Category IV. These assertions require the accumulation of two trace projections over the entire program execution, and complete post-mortem processing. Case 9 is composed of all the previous assertions combined into a single file to yield a monitor that combines multiple assertions on a single execution of the target program.

**Table 1: results for igrep.icn**

| Input Size (text lines) | 4000 | | 16000 | | 64000 | |
|---|---|---|---|---|---|---|
| No monitoring | 0.5 | | 1.6 | | 6.4 | |
| | Events | Time | Events | Time | Events | Time |
| Case 1 | 184208 | 4.1 | 736208 | 16.2 | 2944208 | 1:04.9 |
| Case 2 | 284123 | 4.6 | 1136123 | 18.1 | 4544123 | 1:12.9 |
| Case 3 | 184208 | 3.4 | 736208 | 13.5 | 2944208 | 54.0 |
| Case 4 | 184208 | 3.5 | 736208 | 13.6 | 2944208 | 54.0 |
| Case 5 | 276306 | 6.3 | 1104306 | 28.0 | 4416306 | 2:09.3 |
| Case 6 | 276306 | 6.5 | 1104306 | 28.4 | 4416306 | 2:11.8 |
| Case 7 | 276306 | 6.5 | 1104306 | 29.1 | 4416306 | 2:11.3 |
| Case 8 | 276306 | 6.5 | 1104306 | 29.4 | 4416306 | 2:12.6 |
| Case 9 | 340306 | 45.9 | 1360306 | 3:57.8 | 5440306 | 20:38.6 |

The results depicted in this table allow several observations. Average monitoring speeds on simple assertions in the test environment were in the range of 2-3 million events per minute. Monitoring realistic assertions on real-size programs with real-size input data is quite feasible with this system.

Most assertions impose about one order of magnitude execution slowdown compared with the unmonitored program execution. This is due largely to the coroutine switches between monitor and target program. UFO's primary goal is reducing the effort of writing monitors for experimental research purposes; optimizations to reduce the number of coroutine switches will be desirable so long as they do not sacrifice the flexibility and generality of the current system.

The execution time required by the combination of all assertions (Case 9) is longer than the sums of separate monitor executions. Combined assertion executions have greater memory requirements in the current implementation, because separately collected trace projections compete for available cache and virtual memory resources. Multi-assertion optimizations are not yet implemented in the current UFO compiler.

## 8. Related Work

What follows is a very brief survey to provide the background for the approach advocated in this paper.

### 8.1. Events

The Event Based Behavioral Abstraction (EBBA) method suggested in [5] characterizes the behavior of the entire program in terms of both primitive and composite events. Context conditions involving event attribute values can be used to distinguish events. EBBA defines two higher-level means for modeling system behavior -- clustering and filtering. Clustering is used to express behavior as composite events, i.e. aggregates of previously defined events. Filtering serves to eliminate from consideration events, which are not relevant to the model being investigated. Both event recognition and filtering can be performed at run-time.

An event-based debugger for the C programming language built on top of GDB called Dalek [31] provides a means for describing user-defined events which typically are points within a program execution trace. A target program has to be manually instrumented in order to collect values of event attributes. Composite events can be recognized at run-time as collections of primitive events.

The COCA debugger [9] for the C language uses the GDB debugger for tracing and PROLOG for the execution of debugging queries. It provides a FORMAN-style event grammar for C traces and event patterns based on attributes for event search. The query language is designed around special primitives built into the PROLOG query evaluator.

FORMAN has a more comprehensive modeling approach than EBBA or Dalek, based on an event grammar and a language for expressing computations over execution histories. The event grammar makes FORMAN suitable for automatic source code instrumentation to detect all necessary events. FORMAN's abstraction of event as a time interval provides an appropriate level of granularity for reasoning about behavior, in contrast with the event notion in previous approaches where events are considered point-wise time moments.

Monitoring frameworks such as Dalek and COCA have used GDB to attain a necessary level of abstraction, which in our approach is provided by the Unicon virtual machine. While both approaches yield adequate source-level access and control over the monitored program, the virtual machine approach avoids substantial operating system overhead and offers better performance and scalability to larger programs.

QDB is a query-based debugger for Java programs [23]. QDB focuses on high performance, using a combination of instrumentation, load-time code generation, query optimization, and incremental reevaluation. QDB's portable implementation on top of the Java VM contrasts with Alamo which pervades the Unicon VM. Alamo supports enabling and disabling VM-internal instrumentation on the fly, and does not impose program bytecode instrumentation at compile or load-time. QDB queries focus on OOP relationships such as constraints between object attributes. UFO's event grammar also allows control-structure-oriented event patterns and supports operators such as existential qualifiers not supported by QDB.

### 8.2. Assertion Languages

Assertion (or annotation) languages provide yet another approach to debugging automation. Most approaches are based on Boolean expressions attached to selected points of the target program, like the **assert**() macro in C. In [34] a practical approach to programming with assertions for the C language is advocated, and it is demonstrated that even local assertions associated with particular points within the program may be extremely useful for program debugging

The ANNA [26] annotation language for the Ada target language supports assertions on variable and type declarations. In the TSL [25][33] annotation language for Ada the notion of event is introduced in order to describe the behavior of Tasks. Patterns can be written which involve parameter values of Task entry calls. Assertions are written in Ada itself, using a number of special pre-defined predicates. Assertion-checking is dynamic at run-time, and does not need post-mortem analysis. The RAPIDE project [27] provides an event-based assertion language for software architecture description.

In [5] events are introduced to describe process communication, termination, and connection and detachment of processes to channels. A language of Behavior Expressions (BE) is provided to write assertions about sequences of process interactions. BE is able to describe allowed sequences of events as well as some predicates defined on the values of the variables of processes. Event types are process communication and interactions such as send, receive, terminate, connect, detach. Evaluation of assertions is done at run-time. No composite events are provided.

Another experimental debugging tool is based on trace analysis with respect to assertions in temporal interval logic. This work is presented in [11] where four types of events are introduced: assignment to variables, reaching a label, inter-process communication and process instantiation or termination. Composite events cannot be defined. Temporal Rover is a commercial tool for dynamic analysis based on temporal logics [8].

The DUEL [10] debugging language introduces expressions for C aggregate data exploration, for both assertions and queries.

The FORMAN approach is consistent with many of the ideas introduced in these works.

### 8.3. Program Monitors

PMMS [24] is a high level program monitoring and measuring system. This system works by receiving queries from the user about target programs written in the AP5 high level programming language. PMMS instruments the source code of the target program in order to gather data necessary to answer the posed questions. This data is collected during run time by the monitoring facilities of PMMS and stored in a database for subsequent analysis. Their domain specific query language is similar to FORMAN but tailored for database-style query processing.

JPAX [14], the Java Path Explorer, provides a means to check execution events within a target program based on a user provided specification that is written in Maude, a high level logic language. Like UFO, JPAX supports monitoring based on a virtual machine (JVM in this case). JPAX supports both black box (based on automatic byte-code instrumentation) and white box (based on hand instrumentation) runtime verification.

Dynascope [36] is a system for directing programs written in vanilla C. It consists of a director, an interpreter, and an executor. The director monitors and controls the actions of the target program. The executor is the target program, and the interpreter controls the flow of event streams to and from the director in addition to interpreting the executor. The Dynascope framework provides a way to test and debug programs without altering the source code of the target program.

YODA [22] uses a preprocessor to attach statements to a target Ada program. These statements activate the YODA monitor during program execution. When active, the monitor creates a trace database and a symbol table to aid in debugging. The trace database will contain the program's history regarding variable declaration and use, task synchronization, and change in task status. Prolog queries can be issued by the user in order to confirm or reject hypotheses about program behavior. YODA represents a classical post-mortem trace processing paradigm.

### 8.4. Software Visualization and Debugging

Much of the literature in software visualization is targeted at instruction, including distance learning, e.g. [15][35][6][21]. In such contexts, one can make use of high quality graphic depictions of data such as those in [13]. UFO's aims are more consonant with the "low-fidelity" approach [16]; experience has shown that if graphics are kept simple and incremental algorithms are employed, a very large set of information can be depicted in real time while a program executes.

Software visualization has been used in a debugging context previously, e.g. [29]. Some mainstream debuggers include visualization capabilities, such as DDD [37]. These systems tend to show literal, concrete details about program executions [30]. This works well when the approximate location of interest is known, but does not scale well to large programs or problems where the programmer has no idea where to look. Visualization has been applied to the task of trying to locate bugs [19].

UFO is capable of visualizing multiple views and aspects of code, data, and algorithms; dynamic data is typically presented, with static information (source code, variable scope information, etc.) providing the backdrop in which to interpret what is viewed. Bee/Hive [32] is a multi-component system that presents multiple, coordinated views similar in scope to UFO. UFO exploits a custom architecture and virtual machine support, and focuses on the specification of higher level behavior via a declarative language for reasoning about program behavior.

UFO relies on custom support from a virtual machine implementation. Some software visualization systems have been built on top of debugging support built into the language implementation, including Jinsight [7]. The primary differences between the capabilities provided by the Alamo architecture employed in UFO and those provided by JVMDI include: scope (Alamo's 150+ events versus JVM's ~30), granularity (Alamo's many events per

VM instruction versus JVM's typically one event per instruction).

One system close in spirit to UFO is IBM's PV [20]. PV provides many built-in, low level event types, as well as the ability to hand-instrument higher-level events. UFO does not have PV's access to hardware events, obtained via proprietary cooperation from the AIX operating system, but UFO provides automatic access to higher level semantic events of interest as defined by declarative specifications, and is portable to multiple platforms.

## 9. Conclusions and Future Work

The rising popularity of virtual machine architectures enables dramatic improvements in automatic debugging. These improvements will only occur if debugging is one of the objectives of the virtual machine design, e.g. as in the case of .net [28].

The architecture employed in UFO could be adapted for a broad class of languages such as those supported by the Java VM or the .net VM. Porting to Java or C# would require a new event grammar and either custom extensions to conventional debugging interfaces, or additional bytecode or source code instrumentation. Our approach uniformly represents many types of dynamic analysis activities, such as assertion checking, profiling and performance measurements, the detection of typical errors, and visualization. We have integrated event trace computations into a monitoring architecture based on a virtual machine. Preliminary experiments demonstrate that this architecture is scalable to real-world programs [4].

One of our next steps is to build a repository of formalized knowledge about typical bugs in the form of UFO debugging and visualization rules, and gather experience by applying this collection to additional real-world applications. There remain many optimizations that will improve the monitoring code generated by the UFO compiler; for example, merging common code used by multiple assertions in a single monitor, and generating specialized VMs adjusted to the generated monitor.

## Acknowledgements

This work has been supported in part by U.S. Office of Naval Research Grant # N00014-01-1-0746, and by the National Library of Medicine.

## References

[1] M. Auguston, Program Behavior Model Based on Event Grammar and its Application for Debugging Automation, in the Proceedings of the 2nd International Workshop on Automated and Algorithmic Debugging, AADEBUG'95, Saint-Malo, France, May 22-24, 1995, pp. 277-291.

[2] M. Auguston, A. Gates, M. Lujan, "Defining a Program Behavior Model for Dynamic Analyzers", in the Proceedings of the 9th International Conference on Software Engineering and Knowledge Engineering, SEKE'97, Madrid, Spain, June 1997, pp. 257-262.

[3] M. Auguston, "Lightweight semantics models for program testing and debugging automation", in Proceedings of the 7th Monterey Workshop on "Modeling Software System Structures in a Fast Moving Scenario", Santa Margherita Ligure, Italy, June 13-16, 2000, pp.23-31

[4] Mikhail Auguston, Clinton Jeffery and Scott Underwood. A Framework for Automatic Debugging. Proceedings of the IEEE 17th International Conference on Automated Software Engineering, ASE'02, Edinburgh Scotland, September 2002, pp. 217-222

[5] P. C. Bates, J. C. Wileden, "High-Level Debugging of Distributed Systems: The Behavioral Abstraction Approach", The Journal of Systems and Software 3, 1983, pp. 255-264.

[6] M. Ben-Ari, N. Myller, E. Sutinen, and J. Tarhio. Perspectives on Program Animation with Jeliot. Dagstuhl Seminar on Software Visualization, May 2001, Springer LNCS 2269.

[7] W. DePauw, E. Jensen, N. Mitchell, G. Sevitsky, J. Vlissides, and J. Yang. Visualizing the Execution of Java Programs. Dagstuhl Seminar on Software Visualization, May 2001, Springer LNCS 2269.

[8] D. Drusinsky, The Temporal Rover and the ATG Rover, Lecture Notes in Computer Science, Vol. 1885, pp.323-330, Springer Verlag, 2000.

[9] M. Ducasse, "COCA: An automated debugger for C", in Proceedings of the 1999 International Conference on Software Engineering, ICSE 99, Los Angeles, 1999, pp.504-513.

[10] M. Golan, D. Hanson, "DUEL - A Very High-Level Debugging Language", in Proceedings of the Winter USENIX Technical Conference, San Diego, Jan. 1993.

[11] G. Goldszmidt, S. Katz, S. Yemini, "Interactive Blackbox Debugging for Concurrent Languages", SIGPLAN Notices vol. 24, No. 1, 1989, pp. 271-282.

[12] Ralph E. Griswold and Madge T. Griswold, The Icon Programming Language, 3$^{rd}$ edition. Peer to Peer Communications, San Jose, 1997.

[13] C. Gutwenger, M. Junger, G. Klau, S. Leipert and P. Mutzel. Graph Drawing Algorithm Engineering with AGD. Dagstuhl Seminar on Software Visualization, May 2001, Springer LNCS 2269.

[14] K. Havelund, S. Johnson, and G. Rosu. "Specification and Error Pattern Based Program Monitoring", European Space Agency Workshop on On-Board Autonomy, Noordwijk, Holland, October 2001.

[15] R. R. Henry and K.M. Whaley and B. Forstall. The University of Washington Illustrating Compiler. Proceedings of the ACM SIGPLAN '90 Conference on

[16] C. Hundhausen and S. Douglas. A Language and System for Constructing and Presenting Low Fidelity Algorithm Visualizations. Dagstuhl Seminar on Software Visualization, May 2001, Springer LNCS 2269.

[17] Clinton L. Jeffery, Program Monitoring and Visualization: an Exploratory Approach. Springer, New York, 1999.

[18] Clinton Jeffery, Shamim Mohamed, Ray Pereda, and Robert Parlett, "Programming with Unicon", http://unicon.sourceforge.net.

[19] J.A. Jones, M.J. Harrold, and J.T. Stasko. Visualization for Fault Localization. Proceedings of the ICSE 2001 Workshop on Software Visualization, Toronto, Ontario, Canada, 2001, pp. 71-75.

[20] Doug Kimelman and Bryan Rosenburg and Tova Roth. Strata-Various: Multi-Layer Visualization of Dynamics in Software System Behavior. Proceedings of IEEE Visualization '94.

[21] Igal Kiofman, Ilan Shimshoni and Ayellet Tal. MAVIS: A Multi-Level Algorithm Visualization System within a Collaborative Distance Learning Environment, IEEE Symposium on Human Centric Computing Languages and Environments, September 2002, 216-225.

[22] LeDoux, Carol H. and Parker, D., "Saving Traces for Ada Debugging. Ada in Use", Proceedings of the Ada International Conference, published in ACM Ada Letters, 5(2):97-108, September 1985.

[23] Raimondas Lencevicius, Urs Holzle, and Ambuj K. Singh, "Dynamic Query-Based Debugging of Object-Oriented Programs", Automated Software Engineering, Vol. 10, 2003, 39-74.

[24] Y. Liao, D. Cohen, "A Specificational Approach to High Level Program Monitoring and Measuring", IEEE Transactions On Software Engineering, Vol. 18, No. 11, November 1992, 969 – 978.

[25] D. C. Luckham, D. Bryan, W. Mann, S. Meldal, D. P. Helmbold, "An Introduction to Task Sequencing Language, TSL version 1.5" (Preliminary version), Stanford University, February 1, 1990, pp. 1-68.

[26] D. C. Luckham, S. Sankar, S. Takahashi, "Two-Dimensional Pinpointing: Debugging with Formal Specifications", IEEE Software, January 1991, pp.74-84.

[27] D. Luckham, J. Vera, "An Event-Based Architecture Definition Language", IEEE Transactions on Software Engineering, Vol.21, No. 9, 1995, pp. 717-734.

[28] Microsoft. http://www.microsoft.com/net/

[29] S. Mukherjea and John Stasko. Toward Visual Debugging: Integrating Algorithm Animation Capabilities within a Source Level Debugger. ACM Transactions on Human Computer Interaction 1:3, pp. 215-344, 1994.

[30] Brad A. Myers. Incense: a System for Displaying Data Structures. Computer Graphics. July 1983, pp. 115-125.

[31] R. Olsson, R. Crawford, W. Wilson, "A Dataflow Approach to Event-based Debugging", Software-Practice and Experience, Vol.21(2), February 1991, pp. 19-31.

[32] Reiss, SP. "Bee/Hive: A Software Visualization Back End", in Proceedings of ICSE 2001 Workshop on Software Visualization.

[33] D. Rosenblum, "Specifying Concurrent Systems with TSL", IEEE Software, May 1991, pp.52-61.

[34] D. Rosenblum, "A Practical Approach to Programming with Assertions", IEEE Transactions on Software Engineering, Vol. 21, No 1, January 1995, pp. 19-31.

[35] G. Rossling and T. Naps. Towards Intelligent Tutoring in Algorithm Visualization. Proceedings of the 2nd Program Visualization Workshop, 2002.

[36] R. Sosic, "Dynascope: A Tool for Program Directing", SIGPLAN Notices, Vol. 27, 7, pp. 12-21, Jul, 1992.

[37] T. Zimmermann and A. Zeller. Visualizing Memory Graphs. Dagstuhl Seminar on Software Visualization, May 2001, Springer LNCS 2269.

## Appendix. Syntax for UFO rules

Rules::= ( ( Rule | Within_group ) ';') +

Within_group::=
 'WITHIN' Procedure_name ( ',' Procedure_name ) *
    'DO' ( Rule ';' ) +
    'END_WITHIN'

Rule::= [ Label ':' ]
    [ ('FOREACH' | 'FIND') Pattern
        [ 'FROM' 'PROG_EX' ] ]
    [ ('FOREACH' | 'FIND') Pattern
        [ 'FROM' ('PROG_EX' |
                Metavariable [ '.' (
                    'PREV_PATH' |
                    'FOLLOWING_PATH' )] ) ]
    [ 'SUCH' 'THAT' ] Bool_expr
    [['WHEN' 'SUCCEEDS'] Say_clause + ]
    ['WHEN' 'FAILS'   Say_clause + ]

Implementation constraints:
1) maximum one aggregate operation per assertion, no aggregate operations in assertions with two quantifiers;
2) metavariables within a single assertion should be unique;
3) if FROM is missing, FROM PROG_EX is assumed;
4) aggregate operations can not be nested.

Rule::= [ Label ':' ]
    SHOW ( POINT_PLOT | BAR_CHART )
        TITLE  string
        [ WINDOW_SIZE  width height ]
    X_SIDEBAR
[ SCALING ( linear | exponential | logarithmic ) ]

```
  TICKNUM        ticknum
  MOVING (scroll |rescale | fixed )
  [ INTERVAL_BEGIN interval_begin ]
  [ INTERVAL_END  interval_end ]
  [ TEXT_LABEL string ]
 Y_SIDEBAR
 [ SCALING ( linear | exponential | logarithmic ) ]
  TICKNUM        ticknum
  MOVING (scroll | rescale | fixed )
  [ INTERVAL_BEGIN interval_begin ]
  [ INTERVAL_END   y_interval_end ]
  [ TEXT_LABEL string ]
SOURCE Pattern

( SET
  [ FOREGROUND_COLOR color_name ]
  [ COLOR color_name ]
  [ SHAPE ( square | circle | triangle)]
 X_AXIS
( ( ( ORD | ADD ) '(' (expr | metavariable ) ')' ) | expr )
Y_AXIS
( ( ( ORD | ADD )  '(' ( expr | metavariable )')' ) | expr)
  [( connected ! disconnected )]
) +           ')'

Say_clause ::= 'SAY'
   '(' ( Expression | Metavariable | Aggregate_op ) * ')'

Bool_expr::= Bool_expr1 ( 'OR'  Bool_expr1 )*

Bool_expr1::= Bool_expr2 ( 'AND'  Bool_expr2 )*

Bool_expr2::=  Expr
   [ ( '=' | '==' | '>' | '<' | '>=' | '<=' | '|=' ) Expr ]

Bool_expr2::= 'NOT' Bool_expr2

Bool_expr2::= '(' Bool_expr ')'

Pattern::=  Metavariable ':' Event_type [ '&' Bool_expr ]

Aggregate_op::=  [ ( 'CARD' | 'SUM' ) ] '['  Pattern
        [ 'FROM' ( 'PROG_EX' |
            Metavariable [ '.' ( 'PREV_PATH' |
      'FOLLOWING_PATH' ) ]
                ) ]

  [ 'APPLY' ( Bool_expr | Expression ) ]  ']'

Expression::=   Expr1  (* ( '+' | '-') Expr1 *)

Expr1::=
  Simple_expr  ( ( '*' | 'DIV' | 'MOD' ) Simple_expr )*

Simple_expr::= '-' Simple_expr

Simple_expr::=   integer

Simple_expr::=   Aggregate_op

Simple_expr::=   Metavariable '.' Attribute

Simple_expr::=   string

Simple_expr::=   '(' Expr ')'

Attribute::=
    (    SOURCE_TEXT |
         LINE_NUM |
         COL_NUM |
         TIME_AT_END |
         TIME_AT_BEGIN |
         COUNTER_AT_END |
         COUNTER_AT_BEGIN |
         DURATION |
         VALUE |
         OPERATOR |
         TYPE |
         FAILURE |
         FUNC_NAME |
         ( PARAM_NAMES '[' integer ']' ) |
         FILE_NAME |
         ADDRESS |

      ( VALUE_AT_BEGIN | VALUE_AT_END)
         '(' Unicon_expr ')'
         )

Event_type::=
   (     func_call |
         expr_eval |
         input |
         output |
         variable |
         literal |
         lhp |
         rhp |
         clause |
         iteration |
         test
   )
```